\makeatletter

\newcommand{\Rmnum}[1]{\expandafter\@slowromancap\romannumeral #1@}
\makeatother

\documentclass[12pt,journal,onecolumn]{IEEEtran}

\usepackage{array}
\usepackage{amsmath,graphicx,epsfig,float,subfigure,times,latexsym,verbatim,mathrsfs,amssymb,textcomp,txfonts,color,setspace}
\usepackage{algorithm}
\usepackage{algorithmic}
\usepackage{url}

\IEEEoverridecommandlockouts

\begin{document}

%
\title{A Testbed of Magnetic Induction-based Communication System for Underground Applications}
%
%
%

\author{Xin Tan,~\IEEEmembership{Student Member,~IEEE,}
        Zhi Sun,~\IEEEmembership{Member,~IEEE,}
        Ian~F.~Akyildiz,~\IEEEmembership{Fellow,~IEEE}
\thanks{To appear in IEEE Antennas and Propagation Magazine. Copyright belongs to IEEE.}
\thanks{Xin Tan and Zhi Sun are with the Department of Electrical Engineering, University at Buffalo, the State University of New York, Buffalo, NY 14260, United States. E-mail:
\{xtan3, zhisun\}@buffalo.edu.}
\thanks{Ian~F.~Akyildiz is with the Broadband Wireless Networking Lab at School of Electrical and Computer Engineering, Georgia Institute of Technology, Atlanta, GA 30332, USA. E-mail: ian@ece.gatech.edu}
}

\maketitle
\begin{spacing}{2.0}
\begin{abstract}
Wireless underground sensor networks (WUSNs) can enable many important applications such as intelligent agriculture, pipeline fault diagnosis, mine disaster rescue, concealed border patrol, crude oil exploration, among others. The key challenge to realize WUSNs is the wireless communication in underground environments. Most existing wireless communication systems utilize the dipole antenna to transmit and receive propagating electromagnetic (EM) waves, which do not work well in underground environments due to the very high material absorption loss. The Magnetic Induction (MI) technique provides a promising alternative solution that could address the current problem in underground. Although the MI-based underground communication has been intensively investigated theoretically, to date, seldom effort has been made in developing a testbed for the MI-based underground communication that can validate the theoretical results. In this paper, a testbed of MI-based communication system is designed and implemented in an in-lab underground environment. The testbed realizes and tests not only the original MI mechanism that utilizes single coil but also recent developed techniques that use the MI waveguide and the 3-directional (3D) MI coils. The experiments are conducted in an in-lab underground environment with reconfigurable environmental parameters such as soil composition and water content. This paper provides the principles and guidelines for developing the MI underground communications testbed, which is very complicated and time-consuming due to the new communication mechanism and the new wireless transmission medium.
\end{abstract}

\begin{IEEEkeywords}
Wireless communication, Wireless sensor networks, Transceivers, Magnetic field measurement, Loss measurement, Magnetic induction, In-lab underground environment.
\end{IEEEkeywords}

%
\IEEEpeerreviewmaketitle
\section{Introduction}
\IEEEPARstart{U}{derground} wireless communication is the enabling technology to realize the wireless underground sensor networks (WUSNs), which can be used in a wide variety of novel applications including intelligent agriculture, pipeline fault diagnosis, mine disaster rescue, concealed border patrol, crude oil exploration, among others [1-7].

Underground is more complicated than the terrestrial environment, which contains not only air but also sand, rocks, and water with electrolyte. It is challenging to realize wireless communication in such complex environments. Classic techniques based on electromagnetic (EM) waves are widely used in terrestrial environment. However, those techniques do not work well in underground. First, EM waves experience high levels of attenuation due to absorption by soil, rocks, and water in the underground \cite{Underground}. Second, the electrolyte in underground medium becomes the dominating factor that influence the path loss of EM waves. As a result, the content of water, density and makeup of soil, can change the performance of communication unpredictably since these factors are different in different places and vary dramatically with time. Third, operating frequencies in MHz or lower ranges are necessary to achieve practical transmission range [8,9]. Thus, compared with the communication range, the antenna size will become too large to be deployed underground.

Alternative communication techniques using magnetic induction (MI) provide promising properties that can solve the problems mentioned above. Instead of using propagating EM waves, MI technique utilizes the near field of low frequency EM field to realize the wireless communication. Since this technique is based on the magnetic induction between two coupled coils, it is not influenced by the complicated underground medium since the magnetic permeability of soil is almost the same as that in the air. Moreover, MI communication can effectively transmit and receive wireless signals using a small coil of wire. Hence, the problem of antenna size can be solved.

Theoretical research on MI underground communication is developing fast in recent years but the implementations and validations are seldom developed. Although significant improvement can be achieved by using MI techniques in theory, the theoretical prediction may be biased due to ideal assumptions that are difficult to realize in practical deployment. To this end, a testbed should be designed and implemented to evaluate the MI-based wireless communications in real underground environments. An integrated system, including the signal generator, the transceivers, the reconfigurable underground environment, and observation strategies, needs to be realized.

In this paper, we present the development of a MI-based underground communication testbed as well as the experiments and measurements derived by the developed testbed. In particular, the testbed implements not only the original MI communication mechanism where only single coil is used for each transceiver but also the more advanced MI systems where the MI waveguide [8,10,11] is used to extend communication range and the 3-directional (3D) MI coil \cite{Channel} is used for omnidirectional coverage. The modules in the testbed design include MI coil, signal generation and observation, and underground environment construction. In the MI coil design module, we first use two coupled coils as the original form of MI communication. Then the MI waveguide and 3D MI coils are tested. In the signal generation and observation module, we use universal software radio peripherals (USRP) \cite{USRP} to build an integrated communication system. To construct a reconfigurable underground environment in the lab, 980000 $cm^3$ of sand are poured in a $255$ $cm$ long tank to form the base material of underground medium. A water cycle is built to keep the dynamic balance of water in the tank. Measurements such as path loss, bandwidth, packet error rate (PER) are then taken in the testbed for variance system and environmental configurations. Based on the experiments, the accuracy of the previous theoretical models are evaluated and the comparison with the EM wave-based system is discussed.

The remainder of this paper is organized as follows. The related works are presented in Section \Rmnum{2}. Then the system design, including MI coil, signal generation and observation, and in-lab underground environment, is discussed in Section \Rmnum{3}. After that, we present the system implementation and field experiments, including the operating steps, experimental results, and discussion, in Section \Rmnum{4}. Finally, this paper is concluded in Section \Rmnum{5}.

\section{Related Work}

The concept of WUSNs is first introduced in \cite{Akyildiz_UG_survey_2006}, after which many novel applications are presented based on the WUSNs. In \cite{Structure}, a WUSN is implemented to monitor the underground tunnels to ensure safe working conditions in coal mines. In \cite{Channel}, the WUSNs are deployed during the hydraulic fracturing process in the crude oil extraction, which can provide real-time physical and chemical measurements deep inside oil reservoirs. In [6,7], the WUSNs are utilized for pipeline leakage detection where MI communications are used to connect the sensors along pipelines.

To setup the theoretical fundamentals of the wireless underground communications, the channel models for the propagation of EM waves in underground environments are discussed in [9,13,14]. The EM wave-based underground communication testbed is developed in \cite{Development}. The results from the testbed show that if there is no aboveground device, the EM wave-based communication in pure underground channel encounters prohibitively high path loss.

The MI technique is introduced to wireless underground communication in [8,10,16]. The MI techniques are shown to provide a more reliable underground communication channel. However, MI communication in their original form has a limited communication range due to the high attenuation rate in the near region. The MI waveguide concept is originally developed in [17-22], which is used for artificial delay lines and filters, dielectric mirrors, distributed Bragg reflectors, slow-wave structures in microwave tubes, among others. The channel model for both original MI communication and MI waveguide are maturely developed in recent years [8,10,11,16]. A model of 3D MI communication, is also introduced in \cite{Channel}.

Despite the active theoretical research, the implementations and experimental results on MI underground communications are seldom developed. A few implementation of MI techniques has been done in the areas other than communications. In \cite{Wireless}, MI techniques are used to create a charger along the railway that keeps charging the trains. In [17,24-28], two coupled MI coils are used to transfer the electric power between portable wireless devices. In [17,26], the MI waveguide with strong coupled neighbor coils is tested. None of the above implementation and experiments are used to evaluate the MI-based communications. Moreover, those experiments on MI techniques are taken in the air medium. No experiments have been taken to see the MI communication performance in the complicated underground medium.

To this end, in this paper we develop a testbed for MI underground communication to evaluate the previous theoretical results and provide the experimental platform for future MI communication research.

\section{System design}
\begin{figure}
  \centering
  \includegraphics[width=3.5in]{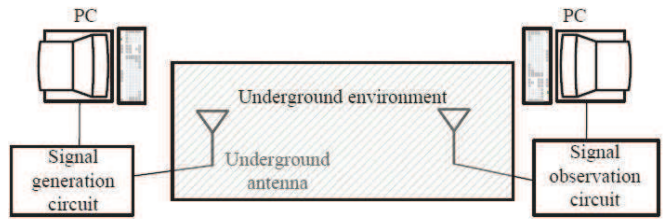}\\
  \caption{The architecture of underground communication testbed.}\label{fig:architecture}
\end{figure}
The underground communication system implies that, the communication occurs entirely using underground propagation medium. To implement and test such an underground-to-underground communication, a testbed should be developed that contains three modules: underground antenna, signal generation and observation, and the in-lab underground environment. Based on these three modules, the architecture of underground communication testbed can be developed as shown in Fig. \ref{fig:architecture}. Two new designed devices, performing as the antennas, are buried in the in-lab underground environment. The signal generation and observation modules, used for signal generation and measurements taking, are connected to the antennas. The initial objective of this testbed is to establish a MI underground communication and evaluate the performance by measuring important characteristics such as path loss, bandwidth, and PER.

\subsection{Underground MI coil design}
\begin{figure}
  \centering
  \subfigure[MI-based wireless communications]{
  \includegraphics[width=3in]{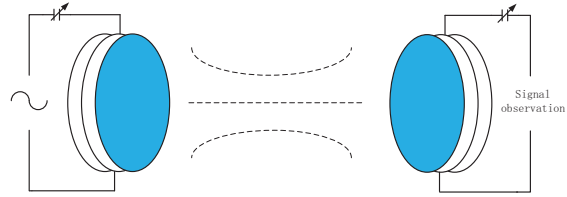}
  \label{fig:p2p_picture}}
  \\

  \subfigure[A photo of MI coil]{
  \includegraphics[width=2in]{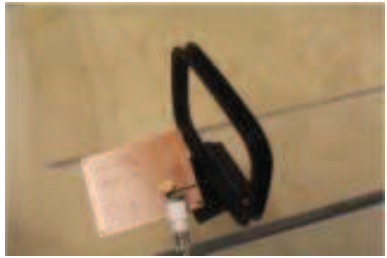}
  \label{fig:p2p_photo}}
  \caption{Wireless communication based on MI}\label{fig:p2p}
\end{figure}

\subsubsection{Original MI communication}
The MI coil plays the most important role in the underground communication system since traditional antennas like electrical dipoles cannot provide a reliable communication in such a complicated propagation medium. As shown in Fig. \ref{fig:p2p_picture}, the transmission and reception are accomplished with the use of a coil of wire. A photo of MI transceivers used for this testbed is shown in Fig. \ref{fig:p2p_photo}. In this testbed, the MI coils are fabricated by 8-turns coils winded on square frames with an edge length of $10$ $cm$. $26$ $AWG$ wire that provide a unit length resistance of $0.1339$ $\Omega/m$ is used for the coils. A variable series capacitor, from $10$ $pF$ to $230$ $pF$, is welded in the circuit to get the resonance.

The ratio of the received power to the transmitted power based on the channel model is developed in \cite{Underground}. Based on this channel model, to increase the channel gain, we can either enlarge the size of the coils or increase the number of turns. However, the size of transceiver gets larger by this way. Other than the coil size and number of turns, the unit length resistance of the loop also have significant influence on the channel gain. Thus, we can use the low resistance wires and circuit to reduce the path loss without increasing the size. To reduce the wire and circuit resistance, we can select high conductivity wires, better connecters and capacitors, and customized printed circuit board (PCB).


Although the MI system in its original form has constant channel condition and relatively longer transmission range than that of the EM wave-based system, its transmission range is still too short for practical applications [7,16]. Moreover, the communication performance of ordinary MI coil is influenced by the intersection angle of two coils so that transceivers can only deployed face to face in a straight line to get the maximum signal strength at the receiver side. To enlarge the communication range and increase the degree of freedom of transceivers deployment, the MI waveguide and 3D MI coils can be used.

\subsubsection{MI waveguide}
\begin{figure}
  \centering
  \subfigure[A view of MI waveguide]{
  \includegraphics[width=3in]{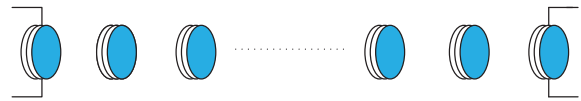}
  \label{fig:waveguide_picture}}\\

  \subfigure[A photo of MI waveguide]{
  \includegraphics[width=2.4in]{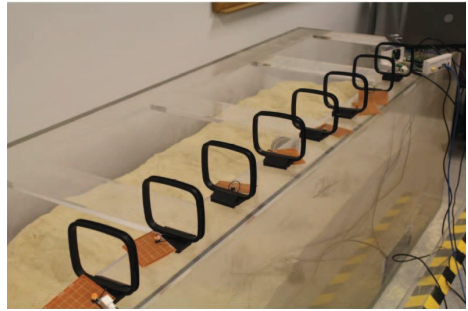}
  \label{fig:waveguide_photo}}
  \caption{MI waveguide based on relays}\label{fig:waveguide}
\end{figure}
The MI waveguide consists of a series of relay coils between two underground transceivers \cite{Deployment}. Different from the relays using EM wave techniques, a MI relay is just a simple coil without any power sources or processing devices. The MI waveguide and the regular waveguide are based on different principles and are suitable for different applications. The MI waveguide is based on the induction coupling between a series or array of independent coils. The communication technique using MI waveguide belongs to the wireless communication although some relay coils are deployed in-between the transceivers. Because of this physical structure, the MI waveguide has a high degree of freedom of deployment and utilization in many harsh environments, such as underground. Fig. \ref{fig:waveguide_picture} gives a view of MI waveguide formed by relays. Since the MI transceivers and relays are coupled by deploying in a straight line, the relays will get the induced currents one by one till the receiver. The signal strength at the receiver side get larger by this way. Fig. \ref{fig:waveguide_photo} is the photo of MI waveguide used in this testbed.

In this testbed, 6 coils are deployed between MI transceivers to perform as the MI relays. Each relay is fabricated by using $26$ $AWG$ wire and a series capacitor, which are the same as the transceivers. The only difference is that relays are close loop circuits without any power input. According to the channel gain of MI waveguide developed in \cite{Underground}, to get a significant increase in signal strength with a communication distance of 2 meters, 6 or more relays are necessary. The density of relays can be reduced if we reduce the unit length resistance for each coil by using high-conductivity wires and PCBs.

\subsubsection{3D MI coil}
\begin{figure}
  \centering
  \subfigure[A view of 3D MI communication]{
  \includegraphics[width=3in]{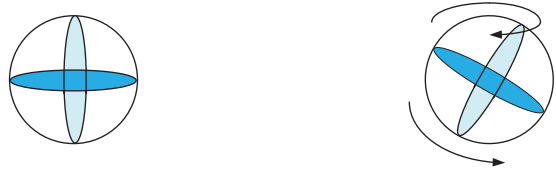}
  \label{fig:3D_picture}}\\

  \subfigure[A photo of 3D MI coil]{
  \includegraphics[width=2.4in]{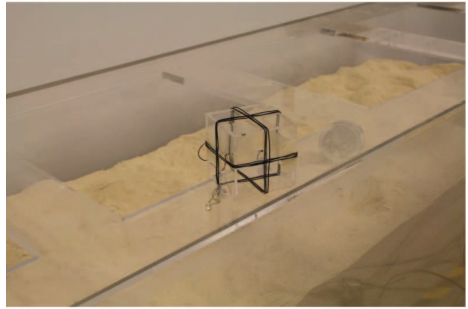}
  \label{fig:3D_photo}}
  \caption{3D MI communication}\label{fig:3D}
\end{figure}
For the MI transceivers, the received signal strength is influenced by the angle between the axes of two coupled coils. However, usually underground MI coils are required to be deployed in a complicated situation that multiple coils are not kept in a straight line. To maintain a high quality of communication in such a complicated situation, an improvement that using multiple-dimensional MI coils is developed in \cite{Channel}. To achieve omnidirectional coverage with minimum number of coils to reduce the system complexity and cost, 3D MI coils designed in \cite{Channel} are used in this testbed. A figure of 3D coil is shown in Fig. \ref{fig:3D} that three independent coils are fabricated and installed vertically on a cube with an edge length of $10$ $cm$. Each of the three coils can form a strong beam along each of the three axes in the Cartesian coordinate. Due to the field distribution pattern of the coil, the orthogonal coils on the same wireless device do not interfere each other since the magnetic flux generated by one coil becomes zero at the other two orthogonal coils. Just like original MI coils, $26$ $AWG$ wire is used for the 3D coil fabrication and each coil is equipped with a series capacitor for finding resonance. At the receiver side, three signals from three independent coils are combined by adding together. Based on the developed channel model, once the parameters of MI coils and communication distance are fixed, how much signal strength we can get depends on the intersection angle between the transmitter and receiver. By using three orthogonal coils, no matter how we change the intersection angle, at least one coil can get enough signal strength. The system is supposed to keep a high quality of communication when rotating the MI coils and changing the intersection angle between them.

\subsection{Signal generation and observation}
\begin{figure}
  \centering
  \includegraphics[width=3.5in]{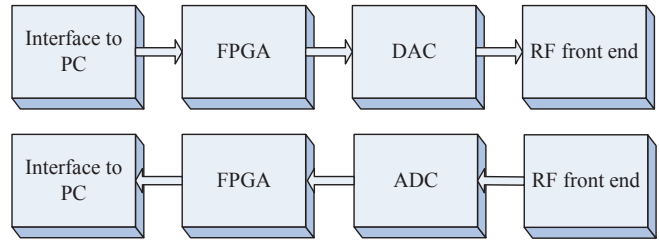}\\
  \caption{Signal generation and observation design}\label{fig:blocks}
\end{figure}
\begin{figure}
  \centering
  \subfigure[A flowgraph of signal generation]{
  \includegraphics[width=3in]{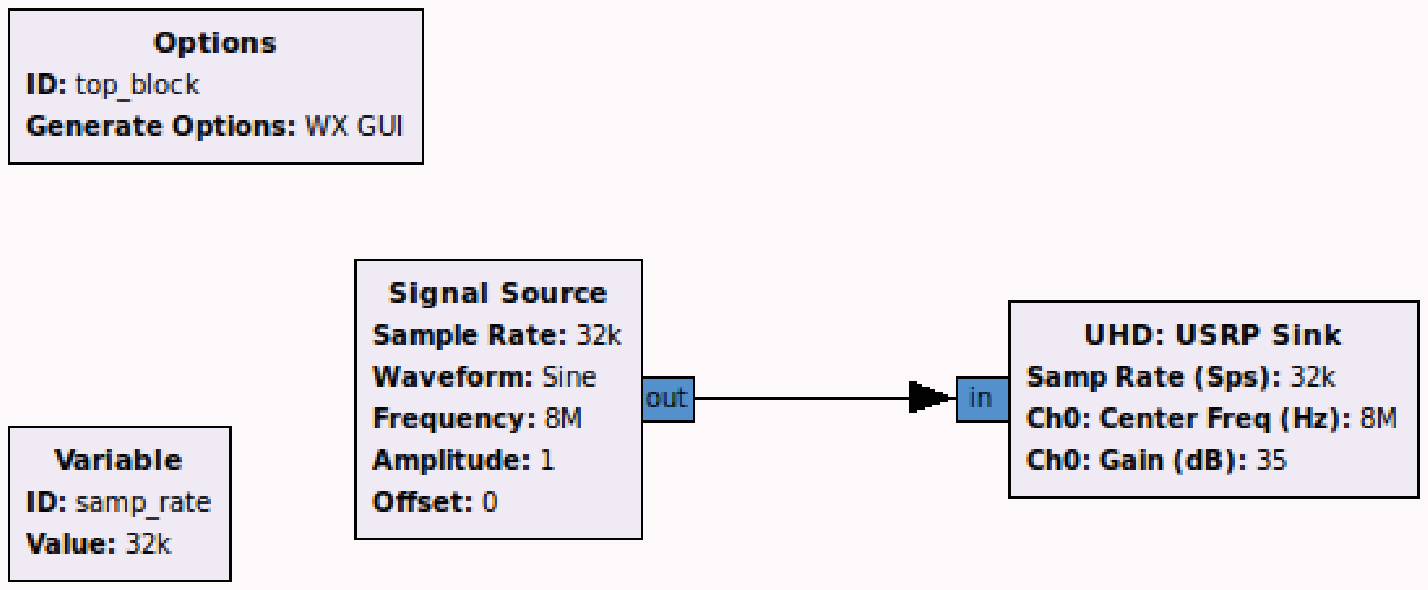}
  \label{fig:GNUgeneration}}
  \subfigure[A flowgraph of signal observation in time domain]{
  \includegraphics[width=3.3in]{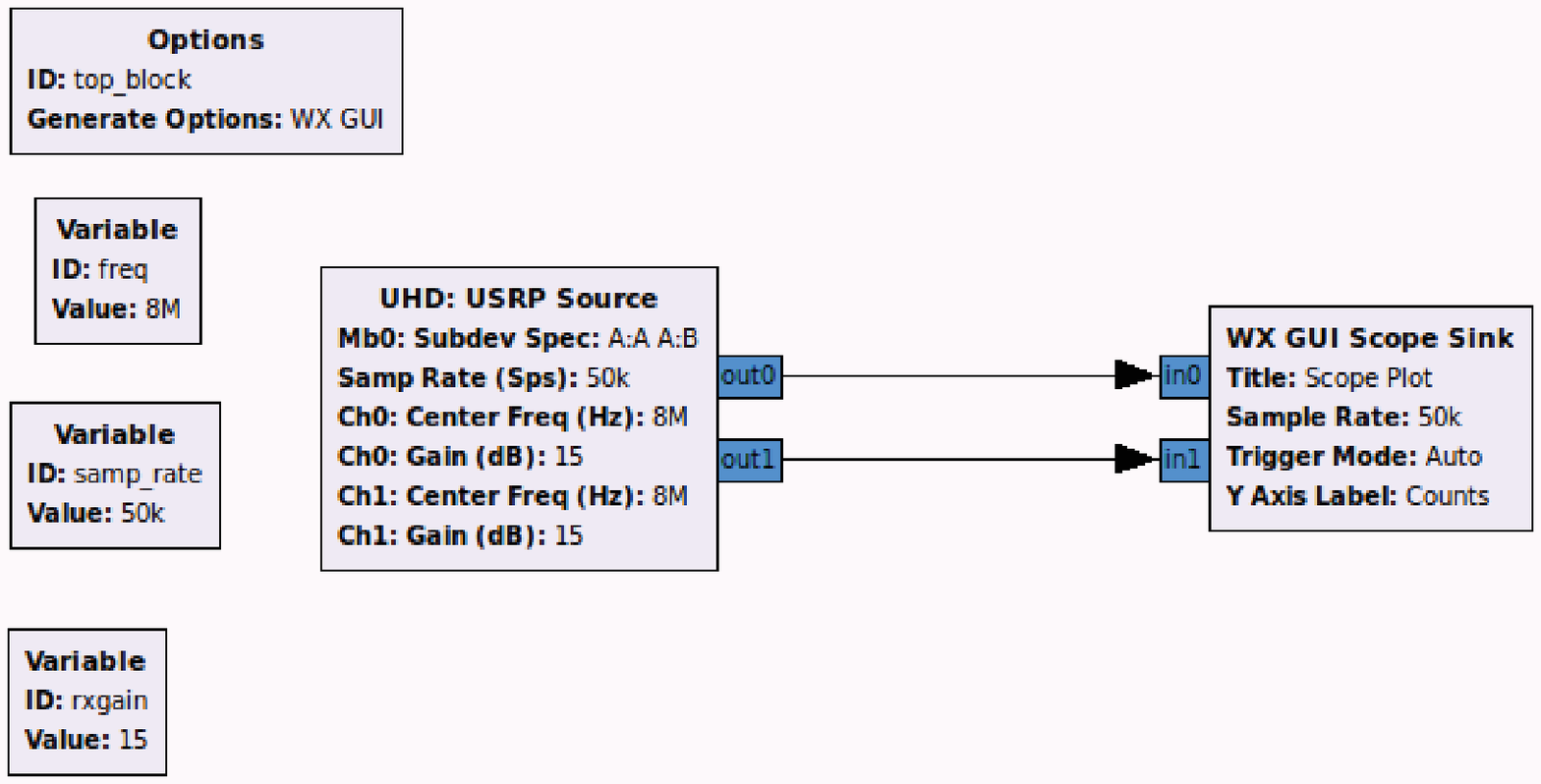}
  \label{fig:GNUtime}}\\
  \subfigure[A flowgraph of signal observation in frequency domain]{
  \includegraphics[width=4in]{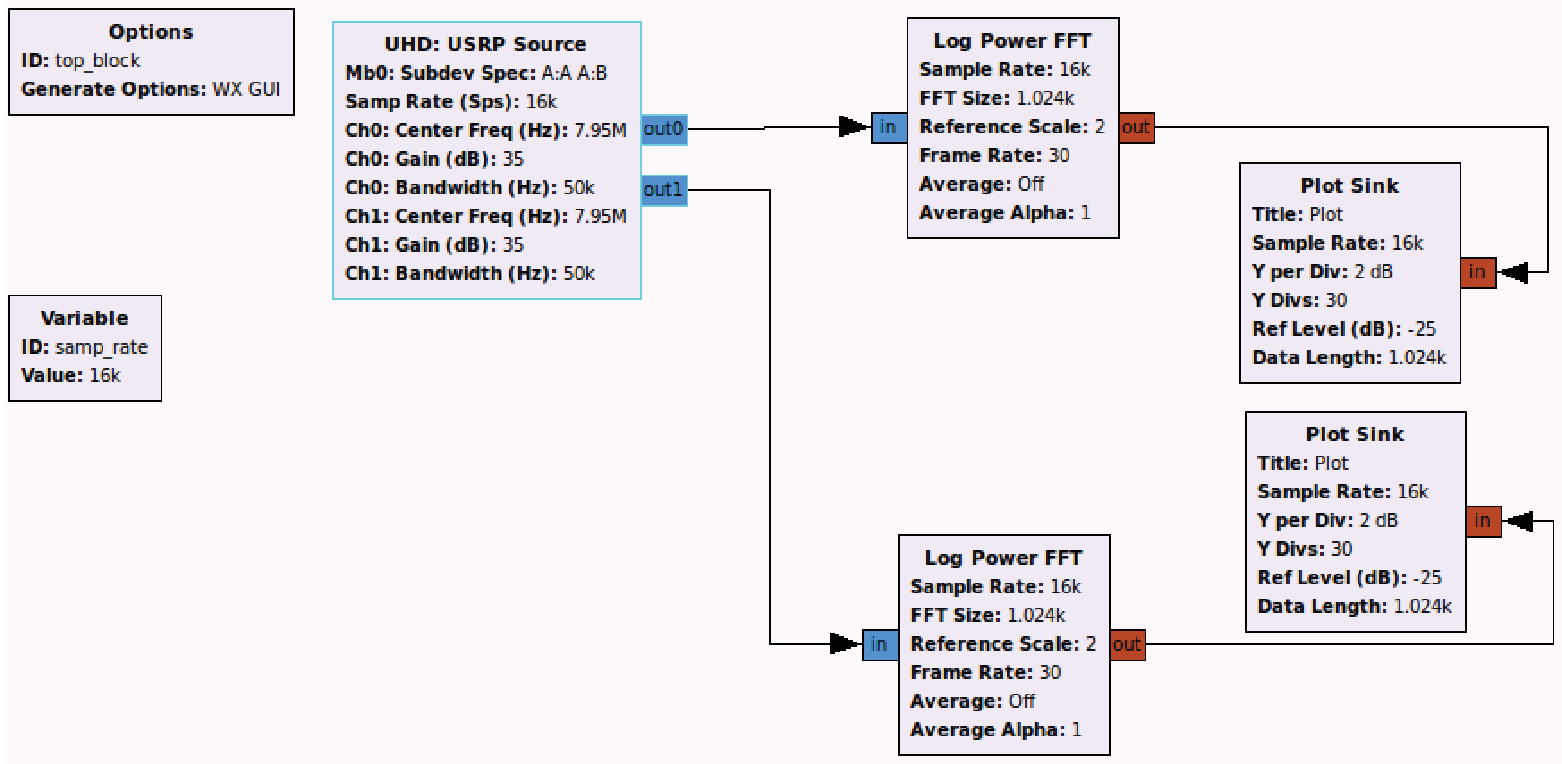}
  \label{fig:GNUfrequency}}
  \caption{Signal generation and observation formed by GRC flowgraphs}\label{fig:GNUflowgraphs}
\end{figure}
\begin{figure}
  \centering
  \includegraphics[width=2.4in]{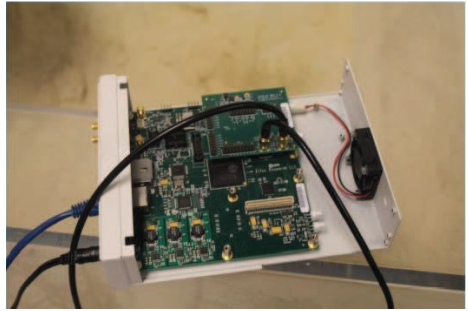}\\
  \caption{A photo of USRP equipped with daughter board}\label{fig:USRP}
\end{figure}
To implement the designed MI coils in underground environment, a signal generator and a method to observe the received signal are needed. A signal generator and an observer can be formed by modules as shown in Fig. \ref{fig:blocks}.

The first module in Fig. \ref{fig:blocks} is the interface to PC, which is connected to a laptop with GNU radio installed. The laptop with GNU radio is used to generate data source and provide the coded information to signal generation circuit though the interface. At the receiver side, GNU radio is also used for signal monitoring. GNU radio is an open source software which achieve the entire signal processing procedure before the transmission. In GNU radio, basic blocks are provided by the inbuilt library. These blocks are written in C++ and are connected using python with the help of swig tools. Blocks can also be created using C++ and added to the library if needed. GNU radio also has a graphical user interface and a GNU radio companion (GRC). GRC is a graphical user interface where different blocks can be connected and the codes can be generated in python according to the connected blocks. These GRC blocks are used in our experiment to send digital signal and read the data at the receiver side.

Fig. \ref{fig:GNUgeneration} shows a flowgraph of signal generation. In this flowgraph, the generated signal is defined by the signal source block. Here we simply use a sinusoidal waveform. The channel gain is defined in a USRP sink block connected to the signal source. The operating frequency, as well as the sampling rate, can also be defined in these GRC blocks. However, this signal generation flowgraph is only used for system correction such as finding the resonance. For further measurements, especially the PER, more complicated signal packets should be sent. Fig. \ref{fig:GNUtime} and Fig. \ref{fig:GNUfrequency} successively show the signal observation in time and frequency domain at the receiver side. In \ref{fig:GNUtime}, a USRP source block is used to get the information from the receiver USRP. A scope sink block is connected directly to show the received signal in time domain. Two independent streams can be read on this scope so that it is convenient to test multiple MI coils at the same time. In \ref{fig:GNUfrequency}, a fast Fourier transformation (FFT) block is added between the USRP source and the scope sink since we need to transform the received signal from time domain to frequency domain and display it in $dBm$ by taking the logarithm. In this flowgraph, we also define two independent channels for signal observation.

The FPGA (field-programmable gate array) module in Fig. \ref{fig:blocks} provides the signal processing and digital up-conversion and interpolation capabilities. After the FPGA, the DAC (digital-to-analog converter) module is used to convert the processed digital information to analog signal that will be input to the transmitting RF front end. At the receiver side, the received signal samples are sent to ADC (analog-to-digital converter) module that decimates and down converts received signal from the receiving RF front end. Further down conversion is done by the FPGA and the samples are sent through the GB ethernet. The FPGA and DAC/ADC blocks can be realized by USRP motherboards \cite{USRP}. In this testbed, USRP N210 is used that equipped with a Xilinx Spartan-3A DSP 3400 FPGA, a 100 MS/s dual ADC, a 400 MS/s dual DAC, and a gigabit ethernet connectivity to stream data to and from host PCs.

The RF front end module in Fig. \ref{fig:blocks} is used to prepare the signal for wireless transmission and input the signal to the MI coil. In the testbed, the daughterboard of USRP is used as the RF front end, which can modulate the output baseband signal to the transmission frequency or convert the received signal back to the baseband. LFTX/LFRX daughterboards are used in this testbed that support transmitting and receiving signals from 0 to 30 MHz. Each LFTX/LFRX daughterboard can support two independent antennas by two antenna connectors TxA/RxA and TxB/RxB. A photo of USRP motherboards equipped with daughterboards is shown in Fig. \ref{fig:USRP}.

\subsection{Underground environment construction}
\begin{figure}
  \centering
  \includegraphics[width=3.5in]{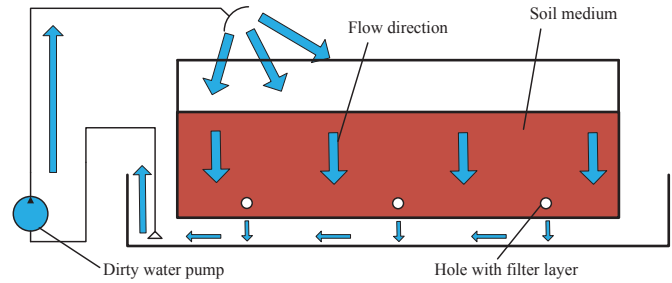}\\
  \caption{A water cycle for the in-lab underground environment}\label{fig:watercycle}
\end{figure}
\begin{figure}
  \centering
  \subfigure[MI coil deployment]{
  \includegraphics[width=2.4in]{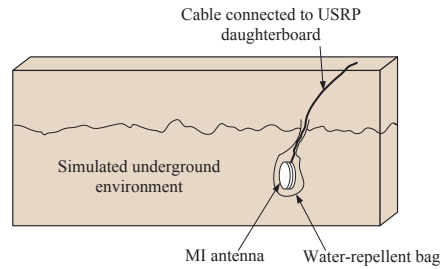}
  \label{fig:underground_picture}}\\

  \subfigure[A photo of the in-lab underground environment]{
  \includegraphics[width=2.4in]{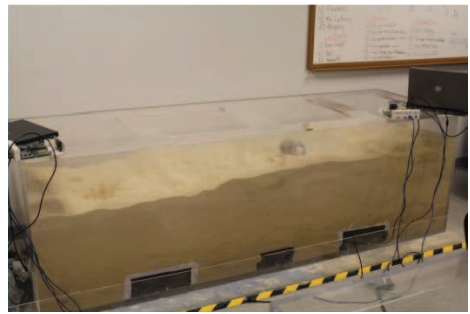}
  \label{fig:underground_photo}}
  \caption{MI coil deployment in the underground environment}\label{fig:underground}
\end{figure}

Since the testbed aims to evaluate the underground MI communications, a underground environment in laboratory is needed. In our implementation, an acrylic tank with a size of $255$ $cm$ $\times$ $76$ $cm$ $\times$ $76$ $cm$ ($length$ $\times$ $width$ $\times$ $height$) is set up on a pedestal. About 980000 $cm^3$ of sand is poured into the tank serving as the base material for the underground environment. The underground soil medium contains a certain concentration of water with electrolyte, which is the dominate factor that can influence the performance of underground communication system. Hence, in the underground environment of the testbed, water should be poured and mixed with the sand. As shown in Fig. \ref{fig:watercycle}, due to the weight and percolation, the water goes down to the bottom of tank and cannot uniform distributed automatically. To address this problem, small holes are opened on the bottom of the tank so that the coming-down water can leak out to the pedestal. A dirty water pump is used to pump the leaking water back to tank from the top. Thus, the water can be circulated inside the tank so that the soil water content is dynamically balanced. In this testbed, water is measured by a measuring cylinder and uniformly mixed with the sand to get a certain water content.

Fig. \ref{fig:underground_picture} shows the method to deploy the MI coils in the underground environment. MI coils are deployed and buried inside tank. Each coil is covered with water-repellent bags to prevent the devices from the water and sand. The MI coil is connected with the USRP daughterboard by well shielded coaxial cable to guarantee that there is no wireless signal radiation from any components other than the coil. A photo of the in-lab underground environment we use is shown in Fig. \ref{fig:underground_photo}.

\section{Experiments and Discussions}

\begin{figure}
  \centering
  \includegraphics[width=3.3in]{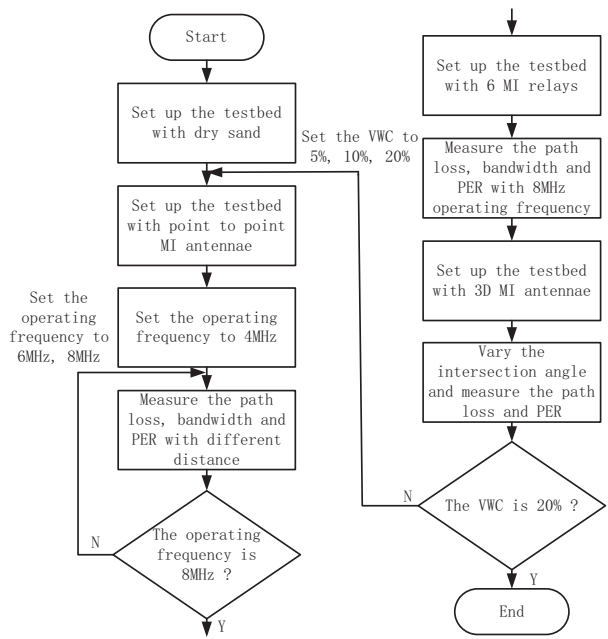}\\
  \caption{A flow chart of operating steps}\label{fig:flowchat}
\end{figure}

To test the performance of MI underground communication in the soil medium with different concentration of water, we vary the volume water content (VWC) from 0\% to 5\%, 10\%, 20\%. Because of the slow evaporation of water, usually the variation of VWC from 0\% to 20\% is irreversible in a short time. For each VWC, we take measurements including path loss, bandwidth, and PER with various communication distances. Each data is measured independently for 10 times.

A flow chart of operating steps is shown in Fig. \ref{fig:flowchat}. First, all the measurements are taken in dry sand. To evaluate the communication performance against the communication range, we vary the distance between transceivers from $20$ $cm$ to $220$ $cm$ and measurements are taken for 10 times for each communication distance. After finishing all the measurements in dry sand, we increase the VWC to 5\%, 10\%, 20\% successively and redo the steps. During the experiment with water inside the tank, a water pump mentioned in Section \Rmnum{3} keeps working to maintain a constant VWC. In this section, the default system configuration includes 8 MHz operating frequency, 8 turns MI coils, and 0\% VWC, if not specified.

\begin{figure}
  \centering
  \includegraphics[width=3.3in]{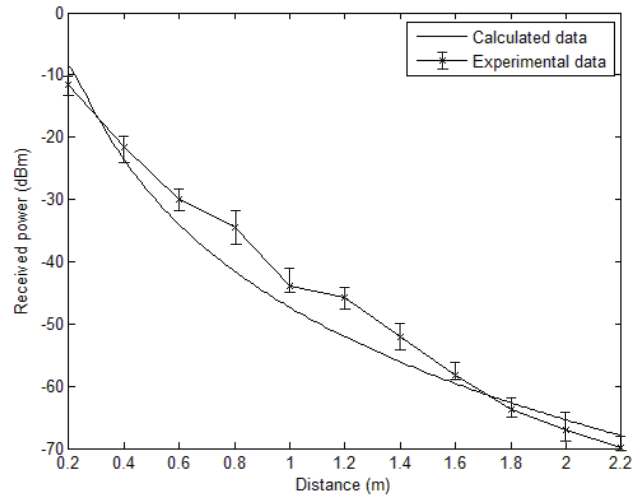}\\
  \caption{Received signal strength of original MI communication}\label{fig:p2p_pathloss}
\end{figure}


Fig. \ref{fig:p2p_pathloss} shows the experimental results derived from the testbed, where the MI communication in the original form (single MI coil) is tested in in 0\% VWC. The x-axis and y-axis respectively show the distance between the transmitter and receiver, the signal strength received at the receiver side. The operating frequency is 8 MHz in this experiment. As shown in this figure, the curve with crosses represents the average received power taken at the receiver side and the smooth curve is the calculated result based on theoretical model. Since the measurements are taken for 10 times at each communication distance, bars on the curve show the maximum and minimum values in each group of data. The experimental result shows a good match with the theoretical calculation. It should be noted that the path loss can be further reduced by using low resistance coil and circuit, which can move the curves in Fig. \ref{fig:p2p_pathloss} upward but the attenuation speed of the curve will not change.

\begin{figure}
  \centering
  \includegraphics[width=3.3in]{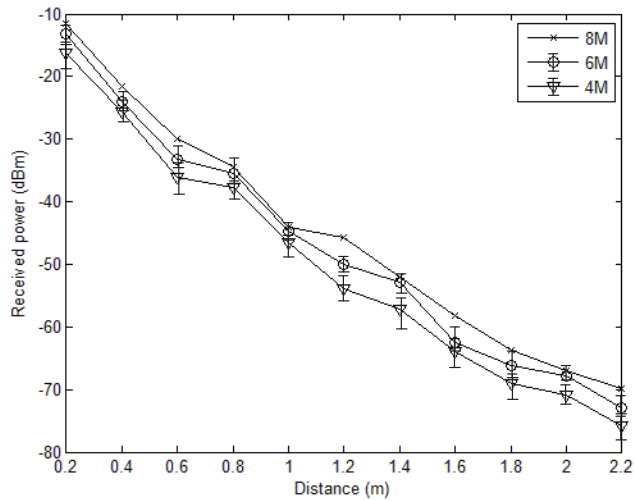}\\
  \caption{Received signal strength for different operating frequency}\label{fig:p2p_frequency}
\end{figure}


Fig. \ref{fig:p2p_frequency} shows the received power with operating frequency variation. Compared to it of 8 MHz, the signal strength is a little lower by using 6 MHz or 4 MHz. As mentioned before, to maximize the signal strength in both transmitter and receiver, a series variable capacitor is welded in each circuit so that the circuit impedance can be minimized by finding the resonance. According to the equivalent RLC circuit developed in \cite{Underground}, since the inductance of the circuit is already determined by the coil fabrication, to cancel the effect of inductance and get the resonance, the value of the capacitor should be changed with frequency variation. However, restricted by the range of series capacitance, the operating frequency can only be taken from 4 MHz to 9 MHz.

\begin{figure}
  \centering
  \includegraphics[width=3.3in]{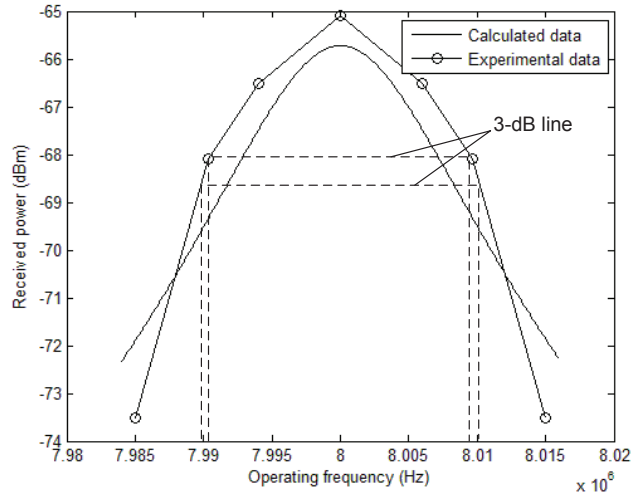}\\
  \caption{Bandwidth of original MI communication}\label{fig:p2p_bandwidth}
\end{figure}

The bandwidth of the MI communication with signal MI coil is shown in Fig. \ref{fig:p2p_bandwidth}. We find the bandwidth of MI system does not change obviously in the distance range that the in-lab underground environment provides. The measurements in Fig. \ref{fig:p2p_bandwidth} are taken by using a communication distance of 2 meters. To find the bandwidth, series capacitors on the MI coil circuites are adjusted for a certain operating frequency (8 MHz) at the beginning. Then the operating frequency is changed to get the decreased signal strength. During the frequency changing, the capacitors are fixed. The bandwidth is then derived by measuring the width of the frequency band where the received signal strength is within a 3 dB range from the peak. From the figure we find that the bandwidth is about 0.02 MHz, which is very close to the calculated result.

\begin{figure}
  \centering
  \includegraphics[width=3.3in]{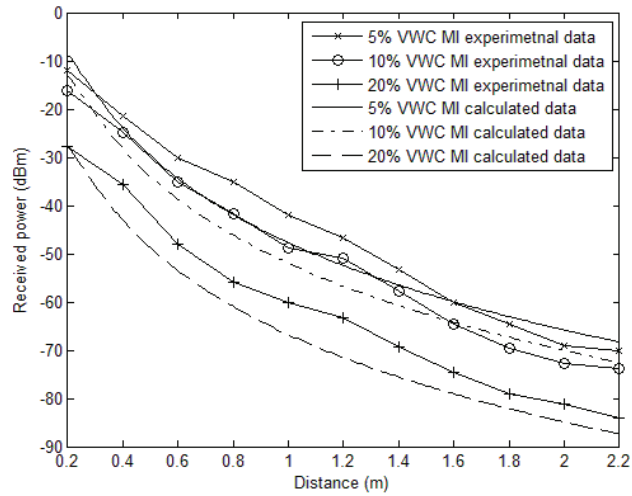}\\
  \caption{Received signal strength in underground environment with different VWC}\label{fig:VWCvstheory}
\end{figure}

\begin{figure}
  \centering
  \includegraphics[width=3.3in]{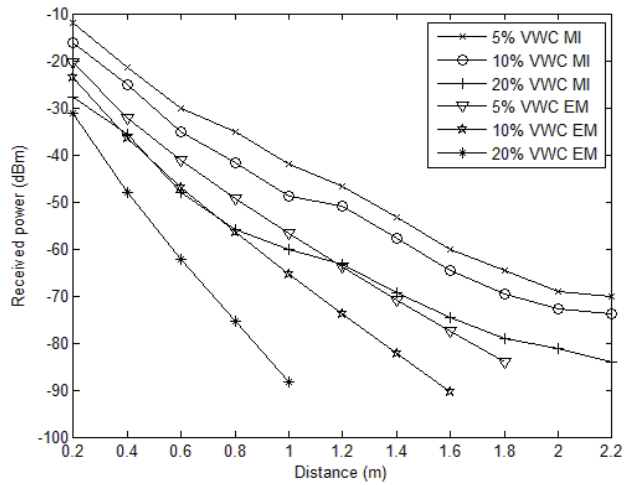}\\
  \caption{A comparison of received signal strength between MI and EM in underground environment}\label{fig:MIvsEM}
\end{figure}

The experimental results of MI communication in underground environments with different water contents are shown in Fig. \ref{fig:VWCvstheory} and Fig. \ref{fig:MIvsEM}. Fig. \ref{fig:VWCvstheory} shows the comparison between the experimental result and calculated result in different VWC underground environments. The communication based on EM waves is also measured to make a comparison in Fig. \ref{fig:MIvsEM}. For the EM power measurements, VERT2450 antennas are used at 2.45 GHz operating frequency. The VERT2450 antennas are dual-band omnidirectional antennas at 3 dBi Gain. The other signal generation and observation modules are the same as used for MI communication. Compared to the communication based on EM waves, even without antenna gain, MI communication shows significantly benefits in underground environment. For 5\% VWC, the received signal strength of using MI communication is about 10 dB higher than it of using EM waves with a communication of 0.2 meters. The difference increases over 20 dB if the communication distance is 1.8 meters. More benefits can be derived if we increase the VWC to 10\% and 20\%. For 20\% VWC, the EM-based communication suffers a high attenuation ratio that the received signal strength is close to -90 dBm with a distance of 1 meter. However, by using MI communication, the signal strength is 30 dB higher in this case.

\begin{figure}
  \centering
  \includegraphics[width=3.3in]{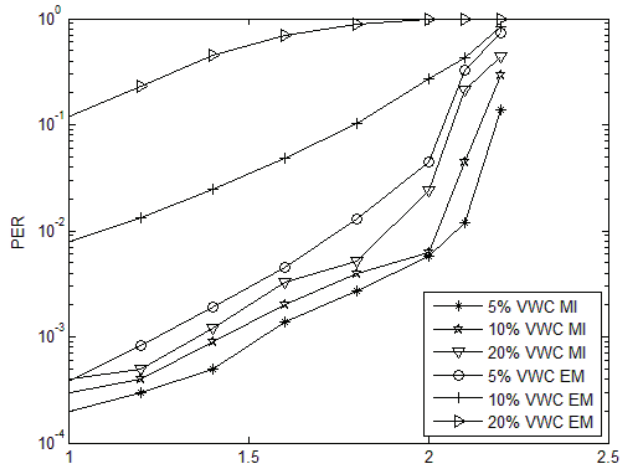}\\
  \caption{A comparison of PER between MI and EM in underground communication}\label{fig:p2p_PER}
\end{figure}

Fig. \ref{fig:p2p_PER} shows the comparison of PER in logarithmic scale between MI communication and EM wave communication in different underground environments. For the MI communication, PER increases with VWC increasing and it becomes obvious after a distance of 2 meters. Thus, the communication range by using this pair of MI coils is around 2 meters. For the communication using EM waves, the PER is much higher than it of using MI communication and it is influenced by the VWC change dramatically. In a 20\% VWC environment, even in a short distance of 1 meter, the PER is over 10\% and a satisfying communication can hardly be established in this case.

\begin{figure}
  \centering
  \includegraphics[width=3.3in]{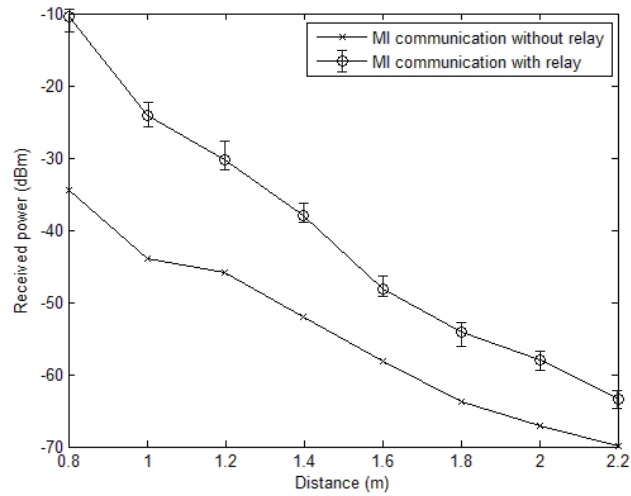}\\
  \caption{A comparison of received signal strength between original MI communication and MI waveguide}\label{fig:p2pvswaveguide}
\end{figure}

\begin{figure}
  \centering
  \includegraphics[width=3.3in]{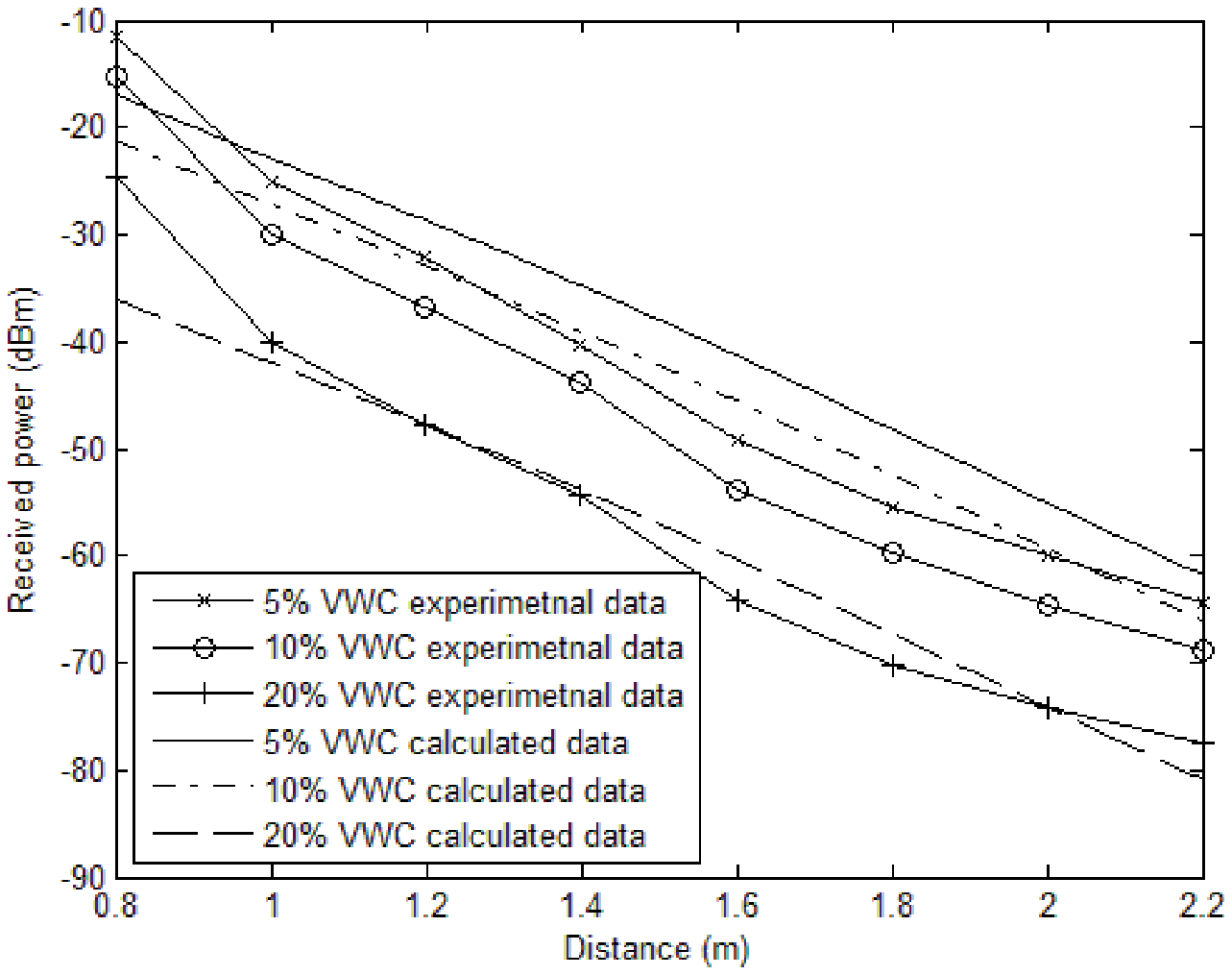}\\
  \caption{MI waveguide in underground environment}\label{fig:waveguide_pathloss}
\end{figure}

\begin{figure}
  \centering
  \includegraphics[width=3in]{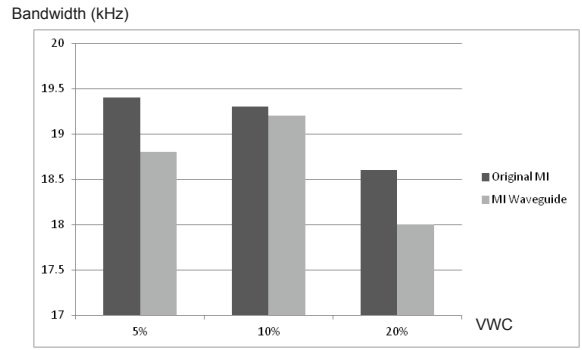}\\
  \caption{A comparison of bandwidth between original MI communication and MI waveguide}\label{fig:p2p_waveguide_bandwidth}
\end{figure}

A comparison between the original MI communication and MI waveguide is shown in Fig.\ref{fig:p2pvswaveguide}. By deploying 6 relays between the transmitter and receiver, the gap between two curves obviously shows the benefits of using MI. Compared to the calculated result, the performance of MI waveguide in different underground environments is shown in Fig. \ref{fig:waveguide_pathloss}. Although the pathloss increases with VWC increasing, the MI waveguide can keep a good performance even in a high VWC of 20\%. A comparison of bandwidth between original MI communication and MI waveguide with 6 relays is shown in Fig. \ref{fig:p2p_waveguide_bandwidth}. 5\%, 10\% and 20\% VWC are successively used to get three comparisons. Compared to the bandwidth of original MI communication, the disadvantage of MI waveguide is not obvious if the MI relays are well fabricated with the same parameters and deployed with equivalent intervals.

Once the received signal strength and bandwidth are determined for both original MI and MI waveguide, the channel capacity of the underground MI communication can be derived. Based on Shannon theorem, the channel capacity can be calculated by substituting the experiment results into the formula $C$$=$$W$$log$$($$1$$+$$\frac{S}{N}$$)$, where $C$ is the channel capacity, $W$ is the bandwidth, and $\frac{S}{N}$ is the signal-to-noise ratio (SNR). For example, since the measured background noise is around -90 dBm, by deploying the MI transceivers with a distance of 2 meters in dry sand and using an operating frequency of 8 MHz, the channel capacity of 448840 bit/s can be derived. The channel capacity of MI waveguide can be calculated in the same way. By deploying the MI transceivers with a distance of 2 meters and 6 relays between them in dry sand, the channel capacity of 608000 bit/s is derived.

\begin{figure}
  \centering
  \includegraphics[width=3.3in]{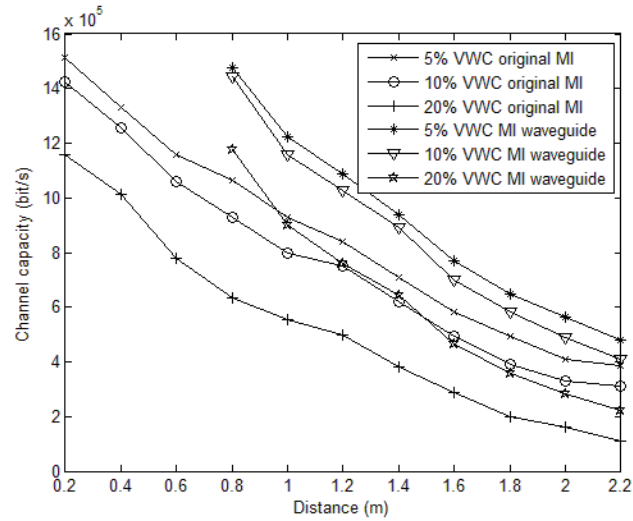}\\
  \caption{Channel capacity of MI communication in different underground environments}\label{fig:capacity}
\end{figure}

Fig. \ref{fig:capacity} shows the channel capacities of MI communication in underground environment with different VWC. For the MI waveguide, we take measurements with a starting distance of 0.8 meters since 6 relays are deployed between transceivers and a necessary distance is kept for two adjacent coils. Since the noise level is a constant, the channel capacity only rely on the bandwidth and received signal strength. For the bandwidth, the difference between using original MI and MI waveguide is not obvious. Moreover, according to the experimental results shown above, the variation of bandwidth is not significant with the changes of VWC and communication distance inside the sand tank. Thus, the received signal strength becomes the dominating factor that influences the channel capacity. As shown in Fig. \ref{fig:capacity}, the performance of MI waveguide is significantly better than the original MI system since MI waveguide dramatically increases the received power. As expected, the channel capacities of both systems become smaller if we increase the VWC to 20\%, which is due to the significant higher path loss in wet soil.

\begin{figure}
  \centering
  \includegraphics[width=3.3in]{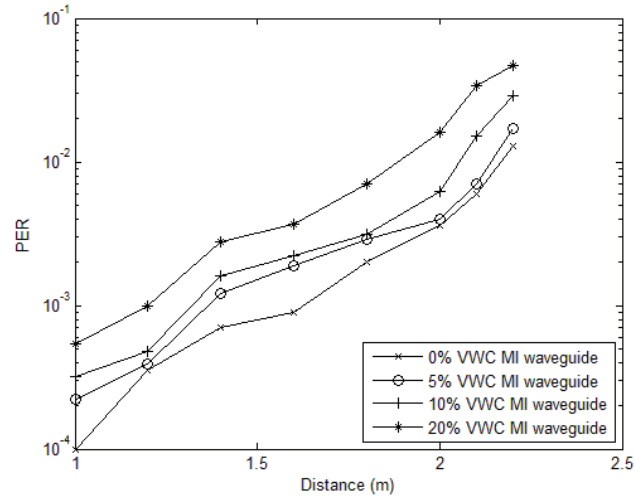}\\
  \caption{PER of MI waveguide in underground environment}\label{fig:waveguide_PER}
\end{figure}

The advantages of MI waveguide are also shown in terms of PER, as shown in Fig. \ref{fig:waveguide_PER}. The PER of MI waveguide keeps in a low level throughout all the experiments. In fact, the communication range of MI waveguide is larger than the length of this testbed, i.e. the length of the tank. Because of the relatively high resistance of wire we used to fabricate the MI coils, non-trivial energy is consumed in these relays. Hence, the experiment results are not the best performance we can have for MI waveguide. Improvements can be achieved by using customized PCB circuit and more efficient experimental supplies. By this way, we can also reduce the relay coil density.

\begin{figure}
  \centering
  \subfigure[1D MI PER]{
  \includegraphics[width=3in]{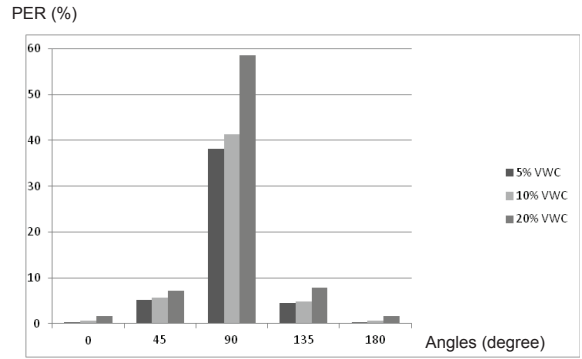}
  \label{fig:1DPER}}\\

  \subfigure[3D MI PER]{
  \includegraphics[width=3in]{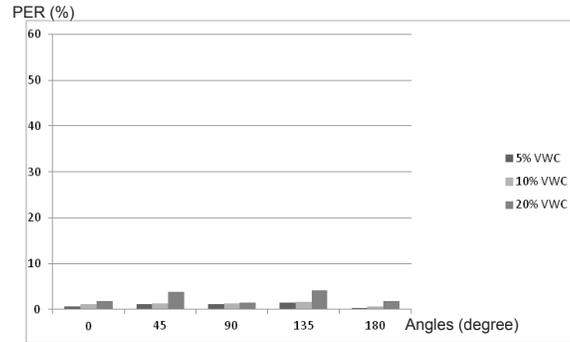}
  \label{fig:3DPER}}
  \caption{A comparison of PER between 1D and 3D MI coils in different VWC underground environments}\label{fig:1Dvs3D}
\end{figure}

As mentioned in Section \Rmnum{3}, 3D MI coil is designed to cover the 3D space in underground environment. The last step of our experiment is to use the testbed to measure the PER for 3D MI coils. A comparison of PER between 1D and 3D MI coils in different underground environments is shown in Fig. \ref{fig:1Dvs3D}. By deploying a 3D MI coil as the receiver with a distance of 2 meters from transmitter and transmitting a series of packets, PER can be measured with intersection angle and VWC changing. We also measure the PER for original MI coil to make a comparison. In this experiment, both original and 3D receiving coils are rotated to change the intersection angle to the transmitting coil. By rotating the receiving coils from 0 degree to 180 degree, we try to find the blind spot of the communication. In Fig. \ref{fig:1DPER}, the original MI coils suffer from a high PER when two coils are deployed orthogonal to each other (90 degree), especially in high VWC environment. However, by using the 3D MI coil, no matter how we change the intersection angle, the PER is kept in a low level throughout the experiments. Even in high VWC environment, a robust communication can still be established.

\section{Conclusion}
In this paper, we develop a testbed of MI-based underground communication system, which validates the feasibility and benefits of using MI for underground applications. First, the experimental measurements are compared to the calculated results based on the developed channel model. Second, according to the experimental results, compared to using EM waves, a significant increasing of signal strength can be achieved by using MI technique, especially in the cases with high VWC. By measuring the PER, we demonstrate that a high-quality communication can be provided by MI technique in underground environment. Third, the testbed implements more advanced MI systems where MI waveguide is used to extend the communication range and the 3D MI coil is used for omnidirectional coverage. For the MI waveguide, a significant improvement of received signal strength can be achieved and the bandwidth is not obviously influenced by the relays. A robust communication is established by using 3D MI coils that we can get a low PER when rotating the receiver.

There is still room for improvements in the future work: instead of hand-made MI coil circuits, we prepare to design PCBs to make the coils printed on boards so that the signal can be received more efficiently and accurately. The PCB-designed circuits will be applied in the fabrication of MI relays and 3D MI coils as well. Second, to achieve the communication distance enlargement and 3D space coverage at the same time, we prepare to design 3D MI relays to let the MI waveguide propagate omnidirectionally. To achieve this, a communication scheme should be developed first in this complicated case.

\section*{Acknowledgment}
This work is based upon work supported by the start-up grant from the State University of New York at Buffalo and the US National Science Foundation (NSF) under Grant No. 1320758.


\ifCLASSOPTIONcaptionsoff
  \newpage
\fi



%

\end{spacing}

\end{document}